\documentstyle[12pt,epsfig]{article}
\begin{document}
\title{Quantum measurements and Paul traps in gravitational backgrounds.}
\author{A. Camacho
\thanks{email: acamacho@nuclear.inin.mx} \\
Department of Physics, \\
Instituto Nacional de Investigaciones Nucleares\\
Apartado Postal 18--1027, M\'exico, D. F., M\'exico.}

\date{}
\maketitle

\begin{abstract} 
In the present work we solve the motion equations of a particle in a Paul trap 
embeded in the gravitational field of a spherically symmetric mass. One of the ideas behind this work concerns the analysis of the effects that the gravity--induced quantum noise, stemming from the bodies in the neighborhood of the Paul trap, could have upon the enhancement of the quantum behavior of this system. This will be done considering a series expansion for the gravitational field of the source, and including in the Hamiltonian of the Paul trap only the first two terms. Higher--order contributions will be introduced as part of the environment of the system, and in consequence will not appear in the Hamiltonian. In other words, we put forward an argument that allows us to differentiate those gravitational degrees of freedom that will appear as an uncontrollable influence on the Paul trap. Along the ideas of the so called restricted path integral formalism, we take into account the continuous monitoring of the position 
of our particle, and in consequence the corresponding propagators and probabilities, associated with the different measurements outputs, are obtained. 
 Afterwards, the differential equation related to a quantum nondemolition variable is posed and solved, i.e., a family of quantum nondemolition parameters is obtained. Finally, a qualitative analysis of the effects on the system, of the gravity--induced environment, will be done.
\end{abstract}

\section{Introduction}

From the very outset of quantum theory (QT) the topic of the so called quantum measurement problem (QMP) [1]
has provoked a deep interest, which has been spawned by the 
unusual and paradoxical results that it predicts.
The quest for a solution of this conceptual difficulty has not only a theoretical 
interest, but also a practical one.
For instance, the detection of very small displacements in the case of gravitational--wave 
antennae [2], or the achievement of high sensitivity in parame\-tric transducers [3] 
(a topic closely related to the design of gravitational radiation detectors), 
requires the analysis of QMP.

Around this topic one of the most interesting ideas in the issue of QMP concerns the so called quantum nondemolition 
measuring processes [4], in which certain class of observables may be measured 
repeatedly with arbitrary precision. 
The fundamental idea behind 
a quantum nondemolition measurement (QNDM) is to monitor a variable such that 
the unavoidable disturbance of the conjugate one does not perturb the time 
evolution of the chosen parameter [3].
Clearly, the dynamical evolution of the corresponding system restricts the 
class of observables that fall within this regime. 

Concerning the experimental possibilities that QNDM could offer in the future, we must state that 
they look very promising [5]. Indeed, for instance, the most important hurdles, that currently
impede the achievement of the corresponding quantum regime in the interaction measuring device--mechanical oscillator, 
have already been identified. In the context of macroscopic mechanical oscillators
the exploration of the quantum behavior could bring, in the near future, breakthrough results [6].

The theoretical works on the topic of QMP comprise already different models that claim 
to solve this long--standing conceptual problem [1, 7]. Here we must add that some of them are equivalent [8], nevertheless, there are also approaches which offer contradictory solutions to QMP [1, 7]. 

Among these models we may find the so called restricted path integral forma\-lism
(RPIF) [9], in which the fundamental point is the restriction, by means of a weight functional, 
of the integration domain of the path integral that gives the corresponding propagator of the analyzed system, when one or more of its parameters 
are subject to a continuous measuring process. 
At this point it is noteworthy to comment that the only parameter of the 
involved measuring device that is contained in RPIF is its experimental resolution. In 
other words, this approach considers no particular measurement scheme, i.e., 
it is a phenomenological, and, in consequence, a very general model. 
Currently, the predictions of RPIF have not been confronted against measurement outputs, 
though at this point we must comment that there are already several works which, in the future, 
could open this possibility [10, 11, 12, 13, 14, 15]. 

In this work we will obtain new theoretical predictions, which could be tested against experimental readouts. 
This will be done as follows. We will consider a particle in a Paul trap [16], 
and assume that the whole system is under the influence of the gravitational field of a spherically symmetric source.
The smallness of the involved distances parameters of a Paul trap, compared with the radius of the Earth, will allow us to take a second order approximation for the gravitational field, and with this simplifiction the corresponding motion equations will be solved. 

It may be argued that the gravitational field plays a negligible role in the dyna\-mics 
of a particle immersed in a Paul trap. Nevertheless, we have already faced some examples in which gravity has, even in the framework of classical physics, surprising effects, as in the Sirius problem [17]. One of the interesting points in connection with the introduction of a gravitational field concerns the fact that if in the case without gravity the classical motion is a periodic one (a fact that appears only if certain conditions upon the experimental parameters are satisfied [18]), then the presence of gravity removes this periodicity from the classical trajectory, i.e., at classical level the presence of the second order term is equivalent to a rescaling of the time independent part of the electric field. 

Concerning the quantum behavior, employing RPIF, a measuring process for one of the coordinates will be introduced. The propagators, 
and their related proba\-bi\-lity densities, associated with the possible measurement 
outputs, will be calculated. Afterwards, the differential equation related to the existence of QNDM variables will be solved, and in this way a family of quantum  nondemolition variables 
will be found. The present work is, in some sense, an extension of previous results [11, 15], in 
which quantum demolition and nondemolition measurements in a Paul trap (without gravitational field) were analyzed.

One of the ideas behind this work is to begin the analysis of the effects of gravity in the emergence of the so called kinematical locality, a characteristic closely related to the phenomenon of decoherence [19]. In this context we face a very common problem, we must define our system, and in consequence the environment surrounding this system. In order to do this we will assume that the gravitational interactions not stemming from the Earth have the same order of magnitude as the third--order term (in a series expansion for the Earth's gravitational field, as a function of the distance to the center of the Paul trap). We may justify this approximation noting that if consider a mass of $m\sim 1$g located at a distance equal to $r\sim 10^{-2}$cm, from the particle caught in the Paul trap, then the Newtonian potential of this mass has the same order of magnitude than the aforementioned third--order term. In this way we may consider the first two terms of the Earth's gravitational potential as part of the system, and higher--order terms (which would then include the gravitational interactions from the rest of the universe) as part of an uncontrollable perturbation, and in consequence as part of the environment of the particle caught in a Paul trap.  
\bigskip
\bigskip

\section{Paul traps and gra\-vi\-tational backgrounds}
\bigskip
\bigskip

\subsection{Motion equations}
\bigskip
\bigskip

In the case of a particle in a Paul trap we have a harmonic oscillator, which 
possesses a frequency equal to $\bar{U} - \bar{V}\cos(\omega t)$, 
being $\bar{U}$, $\bar{V}$, and $\omega$ constants which depend on the electric quadrupole field 
used to trap the particle. 
The motion equations of an electrically charged particle caught in a Paul trap are [16]
(here we assume that ions are injected in the $y$--direction and that there is electric field only along the $x$-- and $z$--coordinates)

\begin{equation}
\ddot{x}(t) + {e\over mr^2}\left[\bar{U} - \bar{V}\cos(\omega t)\right]x(t) = 0,
\end{equation}

\begin{equation}
\ddot{z}(t) - {e\over mr^2}\left[\bar{U} - \bar{V}\cos(\omega t)\right]z(t) = 0,
\end{equation}

\noindent where $e$ represents the electric charge of the particle, $m$ its mass, and $2r$ the distance between the electrodes that constitute part of the 
experimental apparatus. 
As it is already known, Mathieu functions [18] are solutions to this differential 
equation. 

Henceforth we assume that the source of the field is a spherically symmetric body 
with mass $M$ (the Earth), and that the distance between its center and the origin of the la\-bo\-ratory's 
coordinate system is $R$. As mentioned at the end of the last section, higher--order terms of the Earth's gravitational field will be considered as part 
of the environment of the Paul trap. We also su\-ppo\-se that the experimental setup has been arranged such that the vertical direction 
coincides with the $x$--axis. Under these conditions expression (1) becomes

\begin{equation}
\ddot{x}(t) + \left[U - V\cos(\omega t)\right]x(t)  + g = 0,
\end{equation}

\noindent here $U = {e\over mr^2}\bar{U} - 2g/R$, $V = {e\over mr^2}\bar{V}$, and $g = GM/R^2$.
\bigskip
\bigskip

\subsection{Solution to the motion equations}
\bigskip
\bigskip

Let us denote by $X(t)$ any solution to the following equation 

\begin{equation}
\ddot{x}(t) + \left[U - V\cos(\omega t)\right]x(t) = 0.
\end{equation}

At this point we must mention that even though $X(t)$ is a Mathieu function, in
the present situation it is not a solution to the case without gravity. Indeed, $U$ depends, explicitly, upon $g$, i. e., $U = {e\over mr^2}\bar{U} - 2g/R$. 
Only if $2g/R = 0$ (which implies that we have a homogeneous gravitational field) 
is then $X(t)$ a solution to the case in which gravity is absent. Clearly, the introduction of the second--order term implies the rescaling of the time independent part of the electric field.

It is readily checked that the following function is a solution to (3)

\begin{equation}
x(t) = BX(t) + CX(t)\int_{b}^{t}{d\tau\over X^2(\tau)} - 
gX(t)\int_{b}^{t}d\tau\int_{c}^{\tau}{X(t')\over X^2(\tau)}dt',
\end{equation}
\bigskip

\noindent here $B$, $C$, $b$, and $c$ are constants.

The usual solution to the case without gravity, $g = 0$, is recovered imposing 
the condition $C=0$, of course, under this restriction ${e\over mr^2}\bar{U} = U$. 
\bigskip
\bigskip

\section{Continuous measurement of position}
\bigskip
\bigskip

Let us now introduce a continuous measuring process, namely, the $x$ coordinate 
will be monitored. As was mentioned in the first section, in order to consider the 
measuring process we must restrict the integration domain of our path integral [9], but, in a completely equivalent way, we may also introduce in the corresponding path integral a weight functional [20], 
which will contain all the information concerning the measuring process. 
At this point we must choose a particular weight functional, and it will be a gaussian one

\begin{equation} 
\tilde{\omega}_{[a]}= \exp\left\{-{1\over T\Delta a^2}\int_{t'}^{t''}[q - a(t)]^2dt\right\}.
\end{equation} 

We have that $T = t'' - t'$ is the time the measurement lasts, $\Delta a$ denotes 
the re\-so\-lution of the corresponding measuring device, and $a(t)$ is the experimental
output. The reasons for this choice lie on the fact that the results coming from a
Heaveside weight functional [21] and those
coming from a gaussian one [22] coincide up to the order of magnitude.
These last remarks allow us to consider a gaussian weight functional as an approximation of the correct expression.
Hence, it will be supposed that the weight functional of our measuring device has
precisely this gaussian form. We may wonder if this is not an unphysical assumption,
and in favor of this argument we may comment that recently it has been proved that
there are measuring apparatuses which show precisely this kind of behavior [23]. At this point we must underline that our Hamiltonian has an explicit dependence upon time, expression (3), hence concerning the group--theoretical structure associated with time evolution we may not move from a semigroupoid to a semigroup [9]. Additionally it has to be mentioned that a gaussian weight functional has, in this context of group--structure, some drawbacks [9], and that it can only render estimations up to the order of magnitude of the possible effects.

Under these conditions the propagator for this particle reads (the vertical movement 
goes from point $x'$ to point $x''$)

{\setlength\arraycolsep{2pt}\begin{eqnarray}
\hat{U}_{[a]}= \int_{x'}^{x''}d[q]d[p]\exp\left\{{i\over \hbar}
\int_{t'}^{t''}\left({p^2\over 2m} + \tilde{U}q^2 + mgq + {i\hbar\over T\Delta a^2}[q - a]^2\right)dt\right\},
\end{eqnarray}}

\noindent where $\tilde{U} = {m\over 2}[- U + V\cos(\omega t)]$.

This last path integral is gaussian in $p$ and $q$, and therefore, it can be easily calculated [24]

{\setlength\arraycolsep{2pt}\begin{eqnarray}
\hat{U}_{[a]}= \sqrt{m\over2i\pi\hbar x(t')x(t'')\int_{t'}^{t''}x^{-2}(t)dt}
\exp\left\{{i\over\hbar}S_{cl}\right\}
\exp\left\{-{1\over T\Delta a^2}\int_{t'}^{t''}a^2dt\right\}.
\end{eqnarray}

Here $x(t)$ is given by

\begin{equation}
x(t) = BX(t) + CX(t)\int_{b}^{t}{d\tau\over X^2(\tau)} - 
\left[g + 2{i\hbar\over mT\Delta a^2}<a>\right]X(t)\int_{b}^{t}d\tau\int_{c}^{\tau}{X(t')\over X^2(\tau)}dt',
\end{equation}
\bigskip

where $X(t)$ is a solution to the motion equation (also we have that $<a> = {1\over T}\int_{t'}^{t''}a(t)dt$)

\begin{equation}
\ddot{x}(t) + \left[U - V\cos(\omega t) -2{i\hbar\over mT\Delta a^2}\right]x(t) = 0.
\end{equation}

Finally, $S_{cl}$ is the {\it classical} action associated to the motion equation

\begin{equation}
\ddot{x}(t) + \left[U - V\cos(\omega t) -2{i\hbar\over mT\Delta a^2}\right]x(t) + g + 
2{i\hbar\over mT\Delta a^2}<a>= 0.
\end{equation}

If we now rewrite $S_{cl} = S^{(1)}_{cl} + iS^{(2)}_{cl}$, and $x(t) = x^{(1)}(t) + ix^{(2)}(t)$, where $S^{(1)}_{cl}$, $S^{(2)}_{cl}$, $x^{(1)}(t)$, $x^{(2)}(t)$ are all real functions, then the probability density, associated to the measurement output $a(t)$, is

{\setlength\arraycolsep{2pt}\begin{eqnarray}
P_{[a]}= {m\over 2\pi\hbar}\exp\left\{-{2\over T\Delta a^2}\int_{t'}^{t''}a^2dt\right\}\exp\left\{{2\over\hbar}S^{(2)}_{cl}\right\}\nonumber\\
\times\left\{\left[\left(x^{(1)})(t')\right)^2 + \left(x^{(2)}(t')\right)^2\right]\left[\left(x^{(1)}(t'')\right)^2 + \left(x^{(2)}(t'')\right)^2\right]\right\}^{-1/2}\nonumber\\
\times \left\{\left[\int_{t'}^{t''}{\left(x^{(1)}(t)\right)^2 - \left(x^{(2)}(t)\right)^2\over [\left(x^{(1)}(t)\right)^2 + \left(x^{(2)}(t)\right)^2]^2}dt\right]^2 + 4\left[\int_{t'}^{t''}{x^{(1)}(t)x^{(2)}(t)\over [\left(x^{(1)}(t)\right)^2 + \left(x^{(2)}(t)\right)^2]^2}dt\right]^2 \right\}^{-1/2}.
\end{eqnarray}
\bigskip
\bigskip

\section{Continuous nondemolition measurements}
\bigskip
\bigskip

Our Hamiltonian reads

\begin{equation}
H = {p^2\over 2m} + {m\over 2}[U - V\cos(\omega t)]q^2 - mgq.
\end{equation}

In the present case, an observable $A = \rho q + \sigma p$ is called a quantum 
nondemolition variable (QNDV) if the following condition is fulfilled [9]

\begin{equation}
{d\over dt}({\rho\over\sigma}) = {1\over m}({\rho\over\sigma})^2 + m[U - V\cos(\omega t)].
\end{equation}

Let us now consider function $F$ defined as

\begin{equation}
F(t) = -{m\over X(t)}{dX(t)\over dt}.
\end{equation}

It is easily checked that it is a solution to condition (14), 
in other words, ${\rho\over\sigma} = -{m\over X(t)}{dX(t)\over dt}$ satisfies the 
requirement that renders a QNDV. As a matter of fact, we have a family of QNDV, i.e., 
for each function $\sigma:\Re\rightarrow\Re$ we have a QNDV, 
$A(t) = \sigma(t)[p - {m\over X(t)}{dX(t)\over dt}q]$. 
For simplicity let us set $\sigma(t) = 1$, then our QNDV reads

\begin{equation}
A(t) = -{m\over X(t)}{dX(t)\over dt}q + p.
\end{equation}
\bigskip

It is noteworthy to comment that this QNDV depends on the gravitational field. Indeed, 
$X(t)$ depends implicitly upon $g$, i. e., it is a solution to (4). 
Mathematically we have the same function as in the case in which gravity is absent [15], 
nevertheless, in the present case $g$ appears, as a parameter, in $X(T)$, something that does not happen in 
the corres\-pond\-ing expression when $g$ vanishes [15]. 
\bigskip
\bigskip

\section{Conclusions}
\bigskip
\bigskip

We have solved the motion equations for a particle located in a Paul trap, 
when an inhomogeneous gravitational field is present. The probability densities 
that this particle has associated, when its position is continuously monitored, were deduced. 
Afterwards, the differential equation, related to a quantum nondemolition variable, was obtained and solved, i.e., a familiy of quantum nondemolition variables was derived.

As was mentioned in section (2), one of the ideas behind this work comprises the analysis of the effects of the uncontrollable contributions stemming from gravity on a particle caught in a Paul trap. Bearing this in mind we have considered the series expansion of the Earth's gravitational field (as a function of the distance to the center of the experimental device), and introduced the first two terms of this series as part of the Hamiltonian, higher order terms of this field have not been included in it, because they could have the same order of magnitude as the gravitational contributions emerging from other bodies located in the neighborhood of the Paul trap. In other words third or higher--order terms of the gravitational field are defined as part of the environment of our system, and in consequence they do not appear as part of the Hamiltonian. This fact implies that we have put forward an argument that allows us to differentiate those gravitational degrees of freedom that will appear as an uncontrollable influence on the Paul trap.

At this point the question reads: what could be the effects of this gravity--induced environment? For instance, the following term 
in the motion equations reads $3gx^2(t)/R^2$, which will appear in the Hamiltonian as a third 
order term, $-mgx^3(t)/R^2$. Of course, here we have assumed that the gravitational contributions from the rest of the universe (an uncontrollable influence) has this order of magnitude. In the context of Paul traps there are no solutions, yet, to this 
problem. Nevertheless, the example of a harmonic oscillator (considered as a limit of 
a Paul trap, $\bar{V}\rightarrow 0$) could shed some light on the possibilities 
that this new contribution could bring in the case of position monitoring (the QNDM problem has 
a very different context). The current results [25] tell us that a quartic term in the Hamiltonian (not the kind of term that we have here) enhances cla\-ssical behavior (the region in which 
the classical approximation is meaningful suffers an enlargement), and that this enhancement becomes 
more intense as the absolute value of the coefficient of this quartic term increases. If we suppose 
that a third order term could have, qualitatively, the same kind of effect in the case of a 
Paul trap, then we may state that this gravity--induced quantum noise would reduce the spreading of the most probables paths. In other words, the gravitational field of nearby bodies could enhance the classical behavior of the Paul particle. Of course, these last arguments have to be
confronted against the corres\-ponding calculations, a work that up to now has not been done.

The analysis of the possible role that quantum measurements and the gravitational 
field could play in connection with Paul traps has not only an academic interest. 
As it is already known [26], among the most promising candidates for the implementation of quantum computation we may find trapped ions in a Paul trap. In this context, in some of the 
already existing experimental proposals [27], the effects of the coupling of the experimental apparatus with the environment 
are not completely fathomed. The possible relevance of this last point lies on the fact 
that any quantum computer must be isolated from its corresponding environment [28], but the possibility of screening the Paul trap from the gravitational field of its surrounding bodies has to be discarded, in other words these kind of contributions will be always present.
\bigskip
\bigskip
\bigskip

\Large{\bf Acknowledgments}\normalsize
\bigskip

The author would like to thank A. A. Cuevas--Sosa for his 
help. This work was partially supported by CONACYT (M\'exico) Grant No. I35612--E.

\end{document}